\documentclass[%
twocolumn, 
notitlepage, 
aps,
prb, 
10pt, 
superscriptaddress,
floatfix,
letterpaper,
reprint,
bibnotes,
amsmath,amssymb
]{revtex4-2}
 
\usepackage{hyperref}
\usepackage{graphicx}
\usepackage{dcolumn}
\usepackage{bm}
\usepackage[compat=1.1.0]{tikz-feynman}
\usepackage{physics}
\usepackage{lipsum}
\usepackage{booktabs}

\usepackage{changes}
\definechangesauthor[name=Antoine, color=blue]{ag}
\definechangesauthor[name=CNY, color=red]{cnyrev}

\definecolor{linkcolor}{HTML}{399B03}
\definecolor{urlcolor}{HTML}{399B03}
\hypersetup{pdfstartview=FitH, linkcolor=linkcolor,urlcolor=urlcolor,colorlinks=true}
\hypersetup{
    unicode=false,          
    pdftoolbar=true,        
    pdfmenubar=true,        
    pdffitwindow=false,     
    pdfstartview={FitH},    
    pdfauthor={},         
    colorlinks=true,        
    linkcolor=blue,         
    citecolor=red,          
}

\begin{document}

\title{
Electronic correlations and dynamical screening with \emph{ab initio} quantum embedding
}

\author{Chia-Nan Yeh}
\thanks{This author contributed equally to this work.}
\email{cyeh@flatironinstitute.org}
\affiliation{Center for Computational Quantum Physics, Flatiron Institute, Simons Foundation, New York, NY 10010, USA}

\author{Francesco Petocchi}
\thanks{This author contributed equally to this work.}
\email{francesco.petocchi@unifr.ch}
\affiliation{Department of Physics, University of Fribourg, 1700 Fribourg, Switzerland}

\author{Alexander Hampel}
\affiliation{Center for Computational Quantum Physics, Flatiron Institute, Simons Foundation, New York, NY 10010, USA}

\author{Philipp Werner}
\affiliation{Department of Physics, University of Fribourg, 1700 Fribourg, Switzerland}

\author{Olivier Parcollet}
\affiliation{Center for Computational Quantum Physics, Flatiron Institute, Simons Foundation, New York, NY 10010, USA}
\affiliation{Université Paris-Saclay, CNRS, CEA, Institut de Physique Théorique (IPhT), 91191 Gif-sur-Yvette, France}

\author{Antoine Georges}
\affiliation{Collège de France, 11 place Marcelin Berthelot, 75005 Paris, France}
\affiliation{Center for Computational Quantum Physics, Flatiron Institute, Simons Foundation, New York, NY 10010, USA}
\affiliation{CPHT, CNRS, École Polytechnique, Institut Polytechnique de Paris, Route de Saclay, 91128 Palaiseau, France}
\affiliation{Department of Quantum Matter Physics, University of Geneva, 24 Quai Ernest-Ansermet, 1211 Geneva 4, Switzerland}

\author{Miguel A. Morales}
\affiliation{Center for Computational Quantum Physics, Flatiron Institute, Simons Foundation, New York, NY 10010, USA}

\date{\today}

\begin{abstract}
First-principles descriptions of correlated quantum materials require a simultaneous treatment of strong local many-body effects and nonlocal dynamical screening. 
We present an efficient fully self-consistent implementation of $GW$+EDMFT that combines nonlocal correlations at the $GW$ level with a non-perturbative treatment of local correlations within extended dynamical mean-field theory (EDMFT), with a well-defined double-counting correction and a self-consistently determined impurity interaction. 
Self-consistency in both the Green's function and the dynamically screened interaction is essential for describing screening processes across energy scales. 
A key ingredient that makes it computationally feasible is the compression of two-particle correlation functions using interpolative separable density fitting (ISDF).
Applying the scheme to the Mott insulator SrMnO$_3$ and the correlated metal LaNiO$_3$, we show that the self-consistent effective local interaction resolves the overscreening inherent to constrained-RPA approaches. 
For SrMnO$_3$, this stabilizes a Mott-insulating state in quantitative agreement with experiment, while for LaNiO$_3$ local EDMFT corrections yield a quasiparticle mass enhancement consistent with ARPES. 
These results establish fully self-consistent $GW$+EDMFT as a reliable \emph{ab initio} framework for strongly correlated quantum materials.
\end{abstract}

\maketitle

\emph{Introduction}---Accurate \emph{ab initio} descriptions of quantum materials with strong electron correlations remain a central challenge in condensed matter physics. While density functional theory (DFT) offers a favorable balance between accuracy and cost, a universal functional capable of including strong correlation effects is not currently available. A many-body formalism, which also gives access to excited states, is therefore required. 

\begin{figure}[t]
\begin{center}
    \includegraphics[width=\columnwidth]{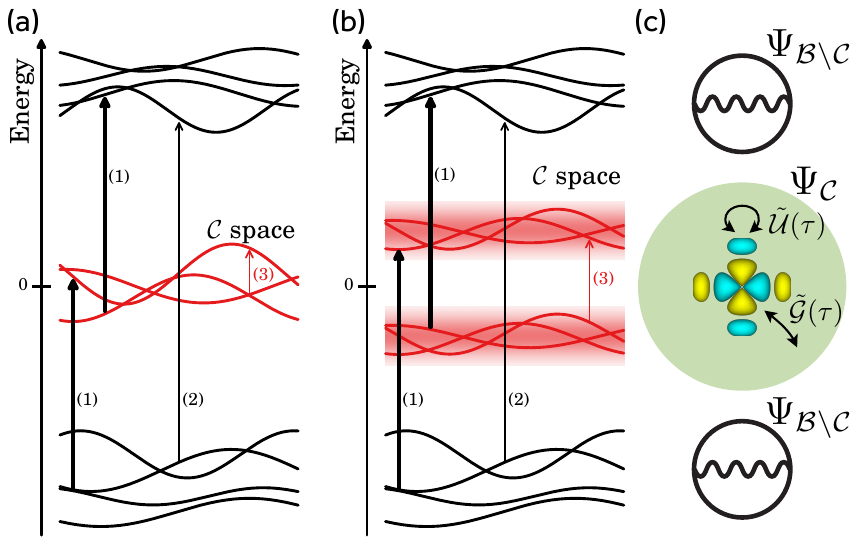}
    \caption{ (a) Different types of screening processes, involving transitions between $\mathcal{B}$ and $\mathcal{C}$ (1), within $\mathcal{B}$ (2) and within $\mathcal{C}$ (3). The cRPA construction removes processes of type (3) from the polarisation, keeping (1) and (2). (b) In a Mott insulator, a gap opens within $\mathcal{C}$. Self-consistent $GW$+EDMFT improves the description of screening processes of type (1) which remedies the overscreening problem of cRPA. (c) Perturbative and non-perturbative contributions to the $\Psi$ functional in $GW$+EDMFT, within $\mathcal{B}$ and $\mathcal{C}$ respectively.}
\label{fig:gw_edmft_cartoon}
\end{center}
\end{figure}

A widely adopted strategy to incorporate many-body effects is \emph{quantum embedding}. The general concept is to isolate, within the full Hilbert space $\mathcal{B}$ of the material, a subspace $\mathcal{C}$ spanned by correlated orbitals to which many-body methods are applied, while the remaining degrees of freedom $\mathcal{B}\backslash\mathcal{C}$ form an environment that is treated at a mean-field or perturbative level (Fig.~\ref{fig:gw_edmft_cartoon}). 

Despite their conceptual simplicity, implementing quantum embedding strategies from first principles poses several major challenges. First, computing effective interactions for $\mathcal{C}$ requires properly accounting for screening of the long-range Coulomb interaction, which involves processes associated with inter-band transitions in the full Hilbert space. In practice, perturbative approaches such as the constrained random phase approximation (cRPA)~\cite{cRPA_Aryasetiawan2004} are often used. cRPA is however known to lead to over-screening in models and materials~\cite{sccRPA_Amadon2014,cRPA_model_benchmark_Hiroshi2015,Honerkamp2018,Loon2021_RPA}. 
The reason is illustrated in Figs.~\ref{fig:gw_edmft_cartoon}a-b: because cRPA is based on an inaccurate description of the correlated states in $\mathcal{C}$, contributions to the polarisation from transitions involving these states are overestimated. This is particularly evident in Mott insulators, in which $\mathcal{C}$ states are pushed away from zero-energy by the gap opening. 
Second, the double-counting (DC) contribution must be defined and removed consistently to ensure a balanced treatment of electronic correlations across Hilbert space partitions, a well-known issue when combining DFT with dynamical mean-field theory (DMFT)~\cite{Haule_prl_2015}. 
An improper treatment of the DC term leads for example to an incorrect relative energy between oxygen-$p$ and metal-$d$ states in transition-metal oxides. 
%
Finally, a self-consistent embedding scheme requires that the perturbative description of the $\mathcal{B}$ space is updated in the presence of many-body corrections arising from $\mathcal{C}$, as done for example in charge self-consistent implementations of DFT+DMFT. 

The combination of the $GW$ approximation~\cite{Hedin65} and extended dynamical mean-field theory (EDMFT)~\cite{EDMFT_Sengupta1995,EDMFT_Si1996,EDMFT_Smith_2000,EDMFT_Chitra_2000}, referred to as $GW$+EDMFT~\cite{GWpEDMFT_Sun2002,GWpEDMFT_Biermann2003}, provides a promising theoretical framework for addressing these challenges. 
The method is formulated as an approximation within the Almbladh free-energy functional framework, where many-body effects are encoded in the correlation functional $\Psi[G,W]$ of the interacting Green's function $G$ and the screened interaction $W$ (Fig.~\ref{fig:gw_edmft_cartoon}c): 
\begin{align}
    \Psi[G,W] \approx \Psi^{GW}_{\mathcal{B}}[G,W] + \Psi^{\mathrm{EDMFT}}_{\mathcal{C}}[G,W] - \Psi^{\mathrm{DC}}_{\mathcal{C}}[G,W]. 
    \label{eq:gw_edmft_psi}
\end{align}
Here $\Psi^{GW}_{\mathcal{B}}$ corresponds to the $GW$ diagram across the entire $\mathcal{B}$ space, incorporating long-range screening and non-local correlations. The \emph{local} correlated subspace $\mathcal{C}$ is treated non-perturbatively with EDMFT, identifying $\Psi^{\mathrm{EDMFT}}_{\mathcal{C}}$ with the Almbladh functional of a self-consistent quantum impurity problem with dynamically screened interactions embedded in an effective non-interacting medium~\cite{GWpEDMFT_Sun2002,GWpEDMFT_Biermann2003} (Fig.~\ref{fig:gw_edmft_cartoon}c). The DC term $\Psi^{\mathrm{DC}}_{\mathcal{C}}$ is defined as the local $GW$ diagram which unambiguously removes contributions common to the first two terms. 

Despite its conceptual appeal, fully self-consistent implementations of $GW$+EDMFT for real materials pose significant numerical challenges, owing to the high computational cost of achieving self-consistency in frequency-dependent two-particle quantities across the full $\mathcal{B}$ space. Consequently, existing fully self-consistent studies have been largely restricted to model systems~\cite{GWpEDMFT_Hubbard_Ayral2012,GWpEDMFT_extended_hubbard_Ayral2013,GWpEDMFT_Si111_model_Hansmann2013,Sakuma2013}, while applications to realistic materials have relied on simplified schemes. These include quasiparticle $GW$ combined with DMFT using static interactions derived from cRPA~\cite{QPGW+DMFT_Tomczak2015,LQSGW_DMFT_Choi2016}, as well as variants that retain frequency-dependent impurity interactions but avoid full self-consistency, such as one-shot~\cite{one_shoft_GWpEDMFT_Tomczak2012,one_shoft_GWpEDMFT_Tomczak2014} and multi-tier schemes~\cite{multitier_GWpEDMFT_Nilsson2017,Mushkaev2024}. As a result, the physical consequences of a fully self-consistent $GW$+EDMFT treatment for real materials have remained largely unexplored, with only very recent work reporting such an implementation with applications to NiO and SrVO$_{3}$~\cite{GWpEDMFT_KANG2025}.


In this article, we introduce an efficient implementation of fully self-consistent \emph{ab initio} $GW$+EDMFT and demonstrate its successful application to two prototypical materials, a Mott insulator ($\mathrm{SrMnO}_{3}$) and a correlated metal ($\mathrm{LaNiO}_{3}$). We show that full self-consistency allows to overcome the issues emphasized above, most notably the overscreening problem, which otherwise prevents a correct description of Mott insulators. Our $GW$+EDMFT implementation is made possible by recent advances in the compression of two-particle correlation functions based on interpolative separable density fitting (ISDF), which enable considerably more efficient self-consistent $GW$ ($\mathrm{sc}GW$) calculations with full frequency dependence~\cite{ISDF_Lu2015,THC-RPA_CNY2023,THCGW_Yeh2024}. Our work demonstrates the viability of $GW$+EDMFT as a first-principle electronic structure framework for materials with strong correlations. 

\emph{Methods}---We introduce a set of Kohn-Sham (KS) Bloch orbitals $\phi^{\bold{k}}_{i}(\bold{r})$ that span the full Hilbert space $\mathcal{B}$. Here ${\bf k}$ is a wave vector in the first Brillouin zone and $i$ is the band index. The correlated subspace $\mathcal{C}$ is defined via a \emph{projector} $C^{\bold{k}}_{ia}$ onto a set of maximally localized Wannier functions (MLWFs), $w_{a}(\bold{r}) = \sum_{\bold{k}}\sum_{i}C^{\bold{k}}_{ia}\phi^{\bold{k}}_{i}(\bold{r})$, with $w_a(\boldsymbol{r})$ centered at the home unit cell. Crucially, an efficient implementation of $\mathrm{sc}GW$ requires a compact basis set for two-particle quantities. Instead of using the product basis set $\rho^{\bold{k}\,\bold{k-q}}_{ij}(\bold{r}) = \phi^{\bold{k}*}_{i}(\bold{r})\phi^{\bold{k-q}}_{j}(\bold{r})$ we adopt the auxiliary basis $\zeta^{\bold{q}}_{\mu}(\bold{r})$ constructed using the ISDF method~\cite{ISDF_Lu2015,THC-RPA_CNY2023,THCGW_Yeh2024}:
\begin{align}
    \rho^{\bold{k}\,\bold{k-q}}_{ij}(\bold{r}) \approx \sum_{\mu} \rho^{\bold{k}\,\bold{k-q}}_{ij}(\bold{r}_{\mu})\zeta^{\bold{q}}_{\mu}(\bold{r})
\end{align}
Here $\bold{r}_{\mu}$ denotes the interpolation points, whose number controls the accuracy of the ISDF representation. At the two-particle level, the mapping between $\mathcal{B}$ and $\mathcal{C}$ is performed using the projector $B^{\bold{q}}_{ab\mu}$, which expands the local Wannier product basis $w_{a}(\bold{r})w_{b}(\bold{r}) = \sum_{\bold{q}}\sum_{\mu}B^{\bold{q}}_{ab\mu}\zeta^{\bold{q}}_{\mu}(\bold{r})$ (for details, see Ref.~\cite{THCGW_Yeh2024}). In the following, a tilde indicates quantities expressed in the MLWF basis of $\mathcal{C}$. 

The Green's function $G$ and screened interaction $W$ in the full Hilbert space, expanded in $\phi^{\bold{k}}_{i}(\bold{r})$ and $\zeta^{\bold{q}}_{\mu}(\bold{r})$ respectively, are related to the self-energy $\Sigma$ and polarisation $\Pi$ by the Dyson equations:
\begin{subequations}\label{eq:dyson_B_space}
    \begin{align}
        &G^{\bold{k}}_{ij}(i\omega_{n}) = \left[ (G_0^{\bold{k}})^{-1}(i\omega_{n}) - \Sigma^{\bold{k}}(i\omega_{n})\right]^{-1}_{ij},\label{eq:dyson_g_B_space}\\
        &W^{\bold{q}}_{\mu\nu}(i\Omega_{n}) = \left[ (V^{\bold{q}})^{-1} - \Pi^{\bold{q}}(i\Omega_{n})\right]^{-1}_{\mu\nu}.\label{eq:dyson_w_B_space}
    \end{align}
\end{subequations}
Here $\omega_{n} = (2n+1)\pi/\beta$ and $\Omega_{n}=2n\pi/\beta$ ($n\in \mathcal{Z}$) are fermionic and bosonic Matsubara frequencies at inverse temperature $\beta$, while $G^{\bold{k}}_{0}$ and $V^{\bold{q}}$ are the non-interacting Green's function and the bare Coulomb interaction. For simplicity, we will omit frequency arguments and basis indices when unambiguous. 

Given the interacting propagators $G$ and $W$, the self-energy and polarization are obtained from the functional derivatives $\Sigma=\delta\Psi/\delta G$ and $\Pi=-2\delta\Psi/\delta W$. Within $GW$+EDMFT, these take the form~\cite{GWpEDMFT_Biermann2003}:
\begin{subequations}\label{eq:sigma_pi_B_space}
    \begin{align}
        &\Sigma^{\bold{k}} = \Sigma_{GW}^{\bold{k}} + C^{\bold{k}\dagger} \left[ \tilde{\Sigma}_{\text{imp}} - \tilde{\Sigma}_{\text{DC}} \right] C^\bold{k}, \label{eq:sigma_B_space}\\
        &\Pi^{\bold{q}} = \Pi_{GW}^{\bold{q}} + B^{\bold{q}\dagger}\left[ \tilde{\Pi}_{\text{imp}} - \tilde{\Pi}_{\text{DC}}\right]B^{\bold{q}}.\label{eq:pi_B_space}
    \end{align}
\end{subequations}
The $GW$ contributions $\Sigma_{GW}^{\bold{k}}$ and $\Pi_{GW}^{\bold{q}}$ are evaluated in the full Hilbert space $\mathcal{B}$ using a cubic-scaling algorithm within the ISDF formalism~\cite{THCGW_Yeh2024}. The corresponding DC terms $\tilde{\Sigma}_{\mathrm{DC}}=-\tilde{G}_{\mathrm{loc}}\tilde{W}_{\mathrm{loc}}$ and $\tilde{\Pi}_{\mathrm{DC}}=\tilde{G}_{\mathrm{loc}}\tilde{G}_{\mathrm{loc}}$ represent the same $GW$ diagrams evaluated using the local projected propagators within the $\mathcal{C}$ space, $\tilde{G}_{\mathrm{loc}}=\sum_{\bold{k}}C^{\bold{k}}G^{\bold{k}}C^{\bold{k}\dagger}$ and $\tilde{W}_{\mathrm{loc}}=\sum_{\bold{q}}B^{\bold{q}}W^{\bold{q}}B^{\bold{q}\dagger}$. 

The impurity self-energy $\tilde{\Sigma}_{\text{imp}}$ and polarization $\tilde{\Pi}_{\text{imp}}$ are obtained by solving an EDMFT impurity problem defined by the fermionic and bosonic Weiss fields $\tilde{\mathcal{G}}^{-1}(i\omega_{n}) = \tilde{G}_{\mathrm{loc}}^{-1}(i\omega_{n}) + \tilde{\Sigma}_{\mathrm{imp}}(i\omega_{n})$ and $\tilde{\mathcal{U}}^{-1}(i\Omega_{n})=\tilde{W}_{\mathrm{loc}}^{-1}(i\Omega_{n}) + \tilde{\Pi}_{\mathrm{imp}}(i\Omega_{n})$ using a numerically exact continuous-time quantum Monte Carlo impurity solver \cite{Werner2010,TRIQS_CTSEG_Kavokine2025}. The EDMFT corrections $\tilde{\Sigma}_{\text{imp}} - \tilde{\Sigma}_{\text{DC}}$ and $\tilde{\Pi}_{\text{imp}} - \tilde{\Pi}_{\text{DC}}$ are then upfolded to the full Hilbert space $\mathcal{B}$ using the projectors $C^{\bold{k}}$ and $B^{\bold{q}}$. 

The lattice Green's function $G^{\bold{k}}$ and screened interaction $W^{\bold{q}}$ are updated using the resulting $\Sigma^{\bold{k}}$ and $\Pi^{\bold{q}}$, which completes one iteration of the self-consistent cycle. The procedure is iterated until the locally projected propagators coincide with the impurity Green's function and screened interaction, $\tilde{G}_{\mathrm{loc}} = \tilde{G}_{\mathrm{imp}}$ and $\tilde{W}_{\mathrm{loc}} = \tilde{W}_{\mathrm{imp}}$. 

\emph{Results}---We assess the performance of self-consistent $GW$+EDMFT by studying two prototypical materials: the Mott insulator SrMnO$_3$ and the correlated metal LaNiO$_3$. SrMnO$_3$ crystallizes in the ideal cubic perovskite structure (space group $Pm\bar{3}m$). In the high-temperature cubic phase, Mn is in a $t_{2g}^3e_g^0$ configuration and the system is a paramagnetic insulator; upon cooling, G-type antiferromagnetic order sets in below $\sim260$~K. Photoemission experiments reveal an insulating gap, even above $T_\text{N\'eel}$, with a valence band dominated by Mn-$t_{2g}$ and O-$2p$ states and a conduction band of predominantly Mn-$e_g$ character~\cite{Saitoh1995,smno_pes_Kang2008,smno_pes_Kim2010}. LaNiO$_3$ is the only rare-earth nickelate of the RNiO$_3$ family that remains metallic down to low temperature. It adopts a nearly cubic perovskite structure with modest octahedral rotations ($R\bar{3}c$). The Ni ions are in a low-spin $t_{2g}^6e_g^1$ configuration in a negative charge-transfer regime, yielding a strongly correlated metal with substantial Ni-$3d$/O-$2p$ hybridization and a narrow band with predominantly $e_g$ character at the Fermi level. Consistent with this picture, experiments report sizeable mass enhancements and proximity to charge and spin instabilities~\cite{lno_nature_comm_Guo2018,Shin2022}. 

All calculations are carried out on a $10\times10\times10$ $k$-mesh at inverse temperature $\beta=15$~eV$^{-1}$. Both compounds are treated in their ideal cubic structures. 
The $\mathcal{B}$ space is spanned by KS orbitals constructed using \texttt{Quantum ESPRESSO}~\cite{QE_Giannozzi2020}, and the $GW$+EDMFT self-consistency loop is initialized from a fully converged $\mathrm{sc}GW$ solution. 
The correlated orbitals and the basis used for spectral comparison are constructed as MLWFs with \texttt{Wannier90}~\cite{Mostofi2014,SupplementalMaterial}. To compare calculations employing different choices of $\mathcal{C}$ for a given compound, we project their spectra onto the same 14-orbital Wannier basis $\mathcal{W}_{14}$, constructed from a large energy window with transition-metal $3d$ and O-$2p$ character. Real-frequency spectral functions are obtained by analytic continuation using the Maximum Entropy (MaxEnt) method~\mbox{\cite{maxent_Bryan1990,maxent_JarrellGubernatis1996}}.

For SrMnO$_3$, we perform two $GW$+EDMFT calculations with different $\mathcal{C}$ spaces constructed from an isolated low-energy manifold spanning $[-2,4]$~eV and having predominantly Mn-$3d$ character. These comprise either the Mn-$t_{2g}$ orbitals ($\mathcal{C}_3$) or the full Mn-$3d$ shell ($\mathcal{C}_5$). For LaNiO$_3$, $\mathcal{C}$ comprises the five atomic-like Ni-$3d$ orbitals contained in $\mathcal{W}_{14}$, constructed using the larger energy window $[-8,5]$~eV and yielding more localized Wannier orbitals. Details of the correlated-subspace constructions, including Wannierization windows and orbital spreads, are provided in the Supplemental Material~\cite{SupplementalMaterial}.

\begin{figure}[t] 
    \begin{center}
    \includegraphics[width=0.45\textwidth]{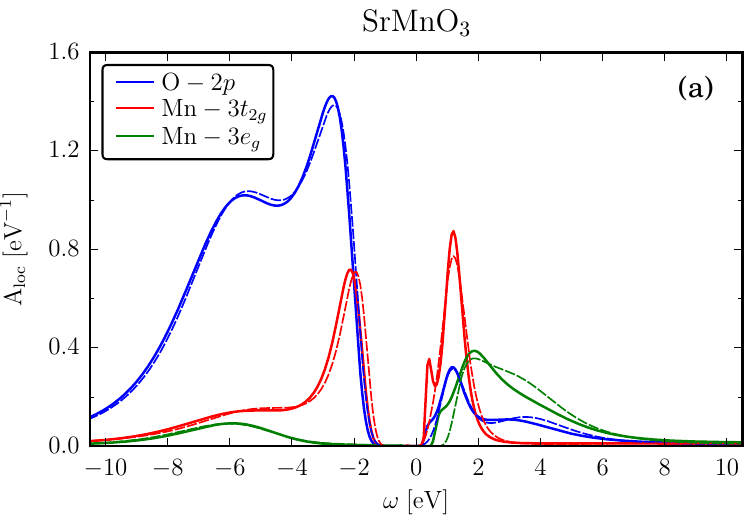}
    \includegraphics[width=0.45\textwidth]{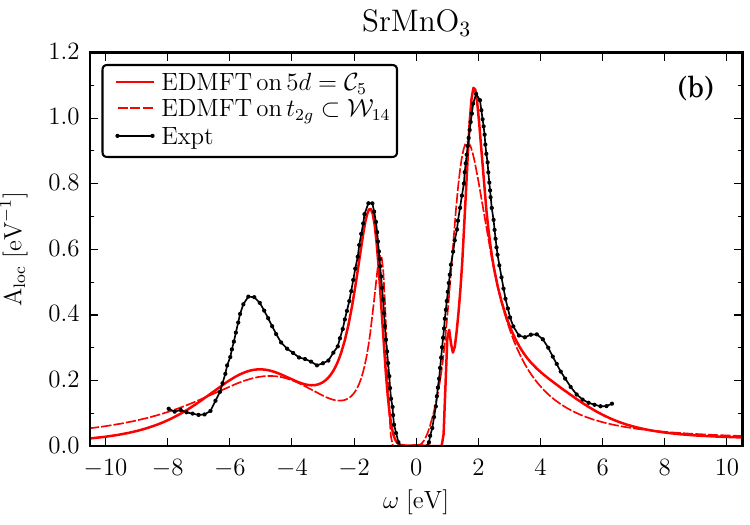}
    \caption{ (a) Local $GW$+EDMFT spectral functions for SrMnO$_3$ projected onto $\mathcal{W}_{14}$, with the EDMFT space given by either the full Mn-3$d$ shell ($\mathcal{C}_5$, thick lines) or the Mn-$t_{2g}$ states ($\mathcal{C}_3$, dashed lines). (b) Mn-$3d$ ($t_{2g}+e_g$) spectral functions compared with PES and XAS data from Ref.~\onlinecite{smno_pes_Kim2010} (black), for $\mathcal{C}_5$ (red thick line) and for a minimal Mn-$t_{2g}$ EDMFT subspace extracted from $\mathcal{W}_{14}$ (red dashed line).}
    \label{fig:smno_Aloc}
    \end{center}
\end{figure}

In Fig.~\ref{fig:smno_Aloc} we report the local spectral functions for SrMnO$_3$. The $GW$+EDMFT method converges to a paramagnetic Mott insulating state, in contrast to the metallic solution obtained at the self-consistent $GW$ level~\cite{SupplementalMaterial}. 
This is a remarkable result, since previous multi-tier calculations, relying on a single-shot $G^0W^0$ description of the high-energy states, failed to obtain an insulating solution in the paramagnetic phase \cite{Petocchi2020,Mushkaev2024}. This is because the DFT $G_0$ on which the calculation of $W_0$ is based has $d$-states at the Fermi level, while these states are in reality located in lower and upper Hubbard bands away from the Fermi level, hence leading to a much weaker screening, as illustrated in Fig.~\ref{fig:gw_edmft_cartoon}a-b. When full self-consistency is achieved, this effect is properly taken into account by $GW$+EDMFT, hence correcting the overscreening problem. This result highlights the ability of $GW$+EDMFT to properly describe spectral properties and screening in Mott insulators. 
In Fig.~\ref{fig:smno_Aloc}b, we compare the $GW$+EDMFT spectrum to photoemission (PES) 
($\omega<0$) and XAS ($\omega>0$) experiments~\cite{smno_pes_Kim2010} (black).
The XAS intensity predominantly reflects the unoccupied 3$d$-derived density of states via $2p$–$3d$ hybridization and hence the onset and overall profile of the XAS spectrum provide a reasonable experimental proxy for the unoccupied part of the local spectral function. 
Our results for the insulating gap and spectra compare quite well to experiments. 
Note that the experimental intensity of the peak at $\sim -5.5$~eV may be quite sensitive to the procedure used in Ref.~\cite{smno_pes_Kim2010} to separate the contribution of the Mn $3d$-states from that of other states.

The close agreement between the $\mathcal{C}_3$ and $\mathcal{C}_5$ results is expected because the Mn-$e_g$ orbitals excluded from $\mathcal{C}_3$ are nearly empty. A similar insensitivity is found for LaNiO$_3$ when the minimal Ni-$e_g$ ($\mathcal{C}_2$) and full Ni-$3d$ ($\mathcal{C}_5$) embeddings constructed from the isolated low-energy window are compared~\cite{SupplementalMaterial}. 
Such insensitivity is, however, material-dependent rather than universal. In general, a physically motivated choice of $\mathcal{C}$ should include all MLWFs with fractional filling, whose exclusion could omit important inter-orbital correlations.

To test the robustness of the embedding scheme, we performed an additional calculation in which the minimal three-orbital Mn-$t_{2g}$ EDMFT subspace is extracted from $\mathcal{W}_{14}$ rather than from the isolated low-energy manifold used to define $\mathcal{C}_3$.
This corresponds to more localized atomic-like orbitals to which non-perturbative many-body effects are applied within EDMFT, as commonly done also in some static implementations of DMFT~\cite{kotliar_rmp_2006}. 
The resulting EDMFT problem and the treatment of the oxygen states are the same as in previous multi-tier studies~\cite{multitier_GWpEDMFT_Nilsson2017,Mushkaev2024}, with the crucial improvement that in the present work the EDMFT self-energy is upfolded to the full space and the latter is treated at the sc$GW$ level. 

The local spectral functions obtained from this scheme, shown in Fig.~\ref{fig:smno_Aloc}b (dashed lines), are in excellent agreement with both the converged results obtained using the other embedding and downfolding choice and with the experimental data. 
The agreement demonstrates internal consistency across the correlated-subspace constructions considered here for SrMnO$_3$.

\begin{figure}[t] 
    \begin{center}
    \includegraphics[width=0.48\textwidth]{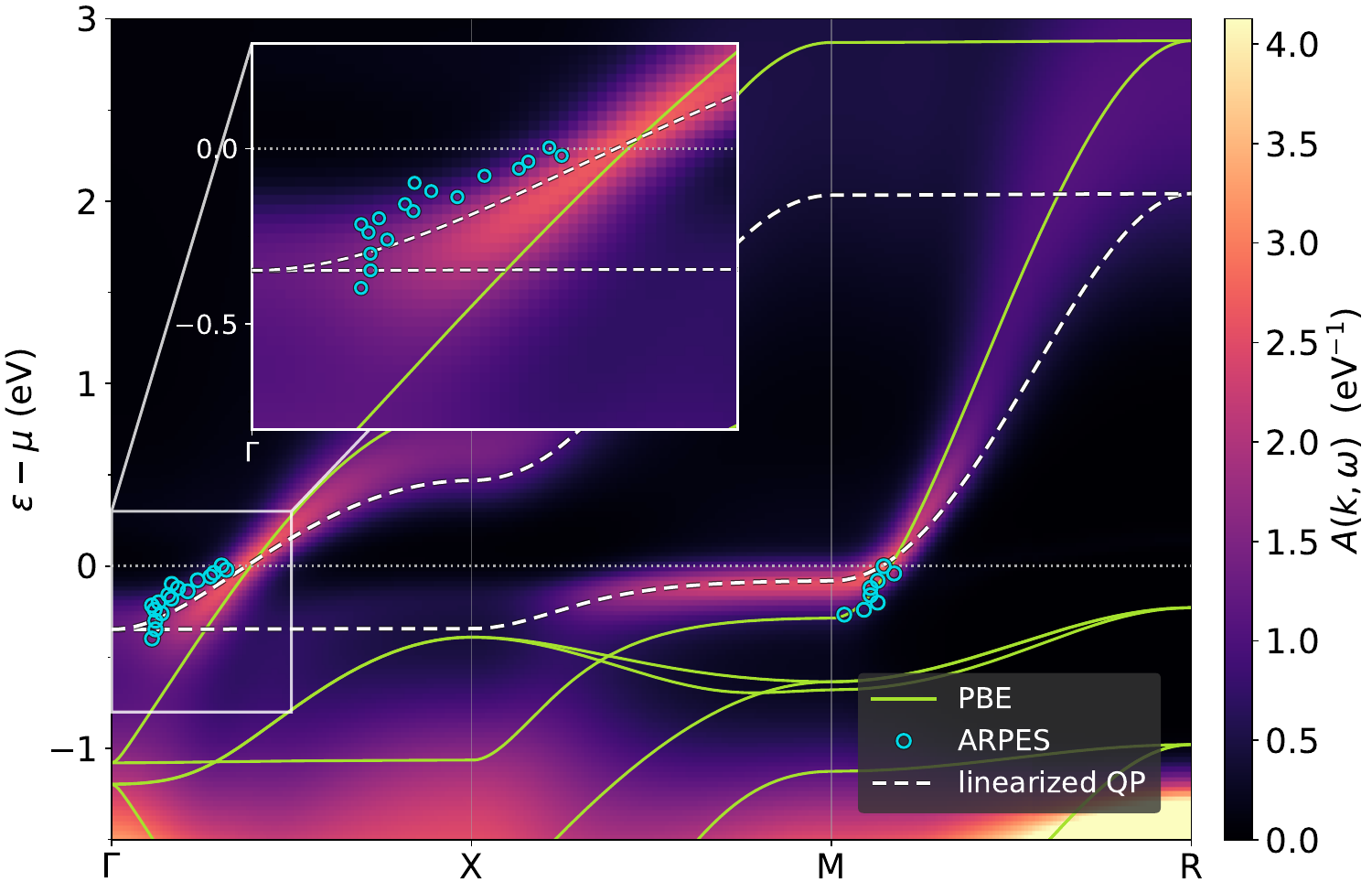}
    \caption{Momentum-resolved spectral function of cubic LaNiO$_3$ obtained from $GW$+EDMFT along the pseudocubic $\boldsymbol{k}$-path. Green lines show the PBE bands, white dashed lines the linearized quasiparticle bands of the $GW$+EDMFT solution, and open circles the ARPES peak positions digitized from Ref.~\onlinecite{Eguchi2009} near $\Gamma$ and $M$. The inset zooms in on the electron pocket near $\Gamma$.}
    \label{fig:lno_k_spectral}
    \end{center}
\end{figure}

Figure~\ref{fig:lno_k_spectral} shows the momentum-resolved spectral function of metallic LaNiO$_3$. Relative to PBE, $GW$+EDMFT substantially narrows the low-energy quasiparticle dispersion, which closely follows the occupied ARPES dispersion along the $\Gamma$--$X$ electron-pocket cut. At the Fermi wave vector of this pocket, we obtain $Z_{\boldsymbol{k}_F}^{-1}\simeq3.2$~\cite{SupplementalMaterial}, consistent with the ARPES electron-band mass enhancement $m^*/m_b\simeq3.3$ reported by Eguchi \emph{et al.}~\cite{Eguchi2009}. Note that a more quantitative comparison would require taking into account the momentum dependence of the self-energy. Near $M$, the agreement is slightly worse, possibly due to missing non-local correlations and crystal-structure effects not captured by the cubic calculation.

\begin{figure}[t]
    \includegraphics[width=\columnwidth]{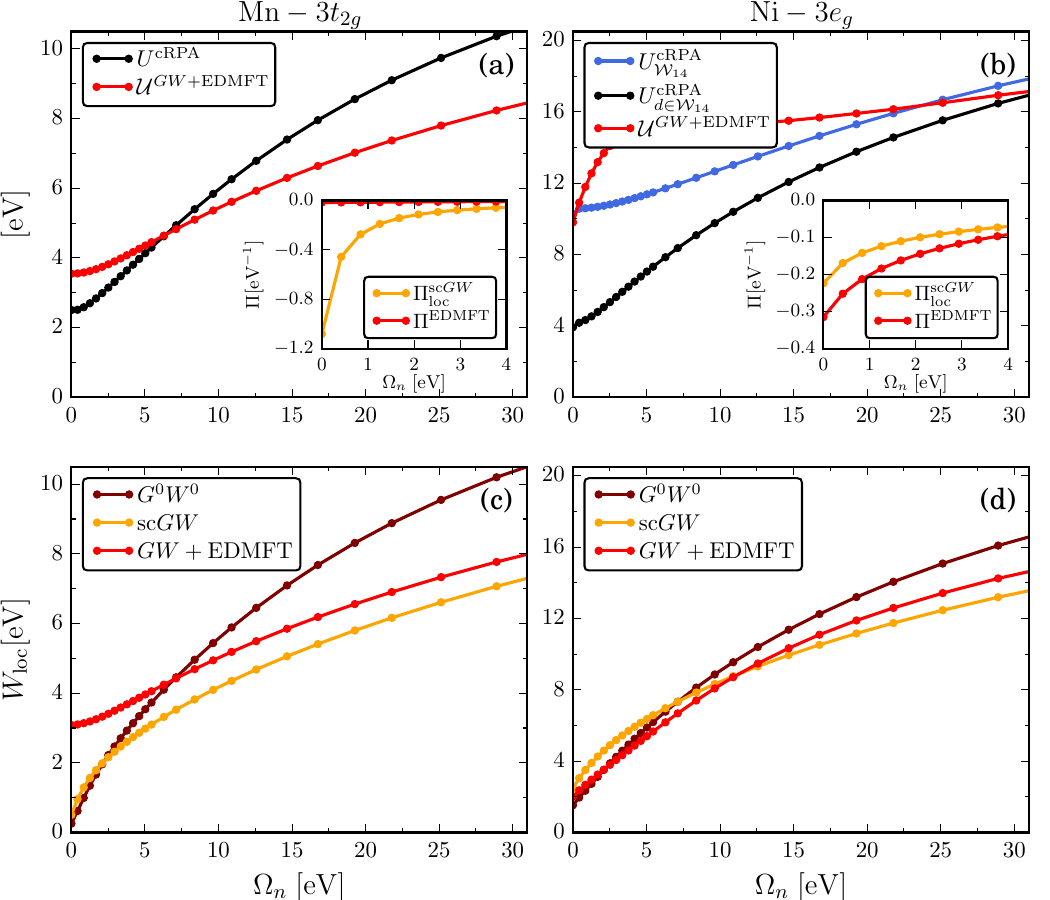}
    \caption{
        Local interactions for Mn-$t_{2g}$ from $\mathcal{C}_{5}$ and Ni-$e_g$ extracted from $\mathcal{W}_{14}$. The top panels compare the local cRPA interaction $U^{\mathrm{cRPA}}$ (black) with the effective local interaction $\mathcal{U}^{GW+\mathrm{EDMFT}}$ of the converged $GW$+EDMFT solution (red). An additional cRPA curve (blue) is shown in (b) as an auxiliary comparison, excluding screening within the full $\mathcal{W}_{14}$ manifold rather than only the Ni-$d$ target space. The insets compare the local EDMFT polarization with the perturbative $\mathrm{sc}GW$ one. Bottom panels: local fully screened interaction $W_{\mathrm{loc}}$ from $G^0W^0$, $\mathrm{sc}GW$, and $GW$+EDMFT for (c) SrMnO$_3$ and (d) LaNiO$_3$. 
    }
    \label{fig:smno_lno_screening}
\end{figure}

The top panels of Fig.~\ref{fig:smno_lno_screening} compare the local cRPA interaction for the Mn-$t_{2g}$ and Ni-$e_g$ MLWFs, computed from the DFT starting point, with the effective local interaction $\mathcal{U}^{GW+\mathrm{EDMFT}}$ of the converged $GW$+EDMFT solution. 
For LaNiO$_3$, strong $d$--$p$ entanglement within $\mathcal{W}_{14}$ makes the identification of particle--hole excitations within the Ni-$d$ target space ambiguous. To address this, we use the spectral cRPA procedure~\mbox{\cite{Kaltak2025_cRPA_spectral,SupplementalMaterial}} to exclude the Ni-$d$ screening channels, yielding $U^{\mathrm{cRPA}}_{d\in\mathcal{W}_{14}}$ (black curve). 
For comparison, the blue curve, $U^{\mathrm{cRPA}}_{\mathcal{W}_{14}}$, shows the cRPA interaction that excludes screening within the entire Ni-$d$+O-$p$ manifold. 
Since it additionally removes the $d$--$p$ and $p$--$p$ screening processes, this construction is not consistent with a low-energy model that contains only Ni-$d$ orbitals.

A first important observation is that, in the low-frequency regime, $\mathcal{U}$ is systematically larger than the cRPA prediction (black curves). This reflects the above-mentioned overscreening of cRPA, which is based on the DFT bandstructure. 
When the self-energy modifies the low-energy electronic structure (e.g., the enhancement of the DFT-underestimated $d$-$p$ splitting~\cite{Carta2026_npj} by the non-local $GW$ self-energy, or the Mott-gap opening in SrMnO$_3$), screening processes are suppressed, resulting in a larger effective interaction.  
A second observation is that, away from the low-frequency regime, the frequency dependence of $\mathcal{U}$ is less pronounced than that of $U^{\mathrm{cRPA}}$.

A relevant check is the comparison between the low-frequency limit of $\mathcal{U}$ and the empirical static interactions commonly used in DFT+DMFT studies. For SrMnO$_3$, $\mathcal{U}(0)\simeq3$--$5$~eV is quantitatively consistent with the values $U=2.5$--$5$~eV used in previous studies, with $J=0.5$--$0.7$~eV~\mbox{\cite{Mravlje2012,Chen2014,Bauernfeind2018,Ricca2019,Long2020}}. For LaNiO$_3$, $GW$+EDMFT yields $\mathcal{U}(0)\simeq9$--$10$~eV, somewhat larger than the values $U=7$--$8$~eV used in previous DFT+DMFT studies, while $J\simeq0.7$--$1$~eV is comparable to the value $J=1$~eV used there~\mbox{\cite{Deng2012,Nowadnick2015}}.
When comparing with cRPA, a closer agreement with $\mathcal{U}(0)$ is obtained only when all screening within the Ni-$d$+O-$p$ manifold is removed, i.e., $U^{\mathrm{cRPA}}_{\mathcal{W}_{14}}(0)\simeq\mathcal{U}(0)$. 
However, this prescription is not consistent with a low-energy model that contains only Ni-$d$ orbitals, and the agreement found here should not be regarded as general. 

The local polarizations shown in the insets of Fig.~\ref{fig:smno_lno_screening} reveal a clear system-dependent contrast between the perturbative $\mathrm{sc}GW$ estimate $\Pi^{\mathrm{sc}GW}_{\mathrm{loc}}$ and the EDMFT impurity polarization $\Pi^{\mathrm{EDMFT}}$. 
For SrMnO$_3$, the RPA-like polarization, indicated by a yellow curve, displays a strongly metallic behavior missing the Mott-insulating ground state, whereas the EDMFT impurity polarization in red is smaller by several orders of magnitude, reflecting the strong suppression of local charge fluctuations. 
In metallic LaNiO$_3$, by contrast, the difference between the two polarizations is more moderate, but interestingly the magnitude of $\Pi^{\mathrm{EDMFT}}$ is larger than the perturbative estimate, indicating enhanced local charge fluctuations captured by $GW$+EDMFT. 

We note that both $W^{G^0W^0}$ (bottom panels) and $U^\text{cRPA}$ (top panels) display a strong frequency dependence over the displayed frequency range. In contrast, self-consistency considerably weakens this frequency dependence, both at the $GW$ and $GW$+EDMFT levels.
At low frequencies, $W^{\mathrm{sc}GW}$ generally exhibits an \emph{increase} relative to $W^{G^0W^0}$, a trend that appears in both materials but is more pronounced for metallic LaNiO$_3$. This behavior indicates that $\mathrm{sc}GW$ underestimates low-energy screening in correlated metals, consistent with its tendency to overestimate band gaps in semiconductors compared to $G^0W^0$. 
Once the EDMFT correction is included, the contrast between metallic LaNiO$_3$ and insulating SrMnO$_3$ becomes evident. In LaNiO$_3$, the $GW$+EDMFT screened interaction smoothly interpolates between the high-energy behavior of $W^{\mathrm{sc}GW}$ and the low-energy behavior of $W^{G^0W^0}$, with a crossover around 5 to 10~eV. 
In contrast, the $GW$+EDMFT bosonic propagator for insulating SrMnO$_3$ differs dramatically from any perturbative result: both $W^{G^0W^0}$ and $W^{\mathrm{sc}GW}$ severely underestimate the magnitude of the screened interaction over a broad frequency range. 
The $GW$+EDMFT screened interaction instead approaches the effective local interaction $\mathcal{U}^{GW+\mathrm{EDMFT}}$ from below. 

These observations lead to two main conclusions: (i) strong local correlations are beyond the reach of perturbative approaches, which hence systematically underestimate the screened interaction in Mott insulators; and (ii) owing to the strongly suppressed impurity polarization, the fully screened interaction $W$ of a Mott insulator remains close to the local effective interaction $\mathcal{U}$. 

\emph{Conclusion}---
We demonstrated that fully self-consistent $GW$+EDMFT~\cite{GWpEDMFT_Sun2002,GWpEDMFT_Biermann2003,GWpEDMFT_KANG2025}, made efficient by ISDF compression of two-particle correlation functions, 
provides 
an accurate 
quantum embedding description of electronic correlations and screening in real materials, with a well-defined double-counting correction and a self-consistently determined impurity interaction. 
Applying this framework to prototypical correlated perovskites, we find that self-consistency stabilizes a Mott-insulating solution in good agreement with photoemission data for cubic SrMnO$_3$, while for metallic LaNiO$_3$ local EDMFT corrections yield a quasiparticle mass enhancement consistent with ARPES. In the Mott insulating system, charge fluctuations are strongly suppressed due to non-perturbative vertex corrections, leading to an effective local interaction that is substantially larger and less structured than predicted by cRPA. In metallic LaNiO$_3$, charge fluctuations are instead enhanced by correlation effects.  
Together, the two materials demonstrate that local vertex corrections reshape screening in a state-dependent manner. 

Our results establish self-consistent $GW$+EDMFT as a practical and reliable \emph{ab initio} framework for materials where
non-local correlations beyond $GW$ remain weak, and motivate applications to systems in which the interplay of screening, covalency, and strong correlations governs metal–insulator transitions, non-Fermi liquid behavior, and other many-body phenomena.

\begin{acknowledgments}

We are grateful to Sangkook Choi, Gabriel Kotliar, Jernej Mravlje, Malte R\"{o}sner and Nils Wentzell for useful discussions. 
F.P. acknowledges the support of the University of Geneva and of the Simons Foundation. 
The Flatiron Institute is a division of the Simons Foundation.

\end{acknowledgments}

\bibliographystyle{apsrev4-2}
\bibliography{refs}

\section*{End Matter}
\appendix
\section*{$GW$+EDMFT benchmark: \NoCaseChange{SrVO$_3$}}
\label{sec:svo}
The two compounds discussed in the main text probe the Mott-insulating and the covalent-metal regimes. As a complementary test in the moderately correlated metal regime, we consider cubic SrVO$_3$, whose spectra are well characterized experimentally and which has served as the canonical $d^{1}$ benchmark for DMFT-type methods~\cite{one_shoft_GWpEDMFT_Tomczak2012,one_shoft_GWpEDMFT_Tomczak2014,Sakuma2013,multitier_GWpEDMFT_Boehnke2016,multitier_GWpEDMFT_Nilsson2017,Mushkaev2024,GWpEDMFT_KANG2025}. As shown below, the fully self-consistent $GW$+EDMFT framework reproduces the main features of the measured spectra.
Experimentally, SrVO$_3$ is a correlated metal with a mass enhancement $m^*/m_b\simeq2$ 
deduced
from both ARPES~\cite{svo_arpes_Yoshida2005,svo_arpes_Yoshida2010} and specific heat~\cite{svo_specific_heat_Inoue1998} measurements, an occupied $t_{2g}$ band bottom at $\simeq-0.5$~eV~\cite{svo_arpes_Yoshida2005,svo_arpes_Takizawa2009}, an incoherent photoemission feature around $-1.5$~eV~\cite{svo_pes_ipes_Morikawa1995,svo_pes_Sekiyama2004,svo_arpes_Aizaki2012}, and a broad unoccupied feature at $2.5$--$3$~eV in inverse photoemission and XAS~\cite{svo_pes_ipes_Morikawa1995,svo_pes_xas_Inoue1994}.
The calculations use the same $k$-mesh and temperature as in the main text. The EDMFT subspace $\mathcal{C}$ comprises the full V-$3d$ shell, constructed from a frontier energy window as in the $\mathcal{C}_5$ setup for SrMnO$_3$~\cite{SupplementalMaterial}. 

\begin{figure}[t]
    \begin{center}
    \includegraphics[width=\columnwidth]{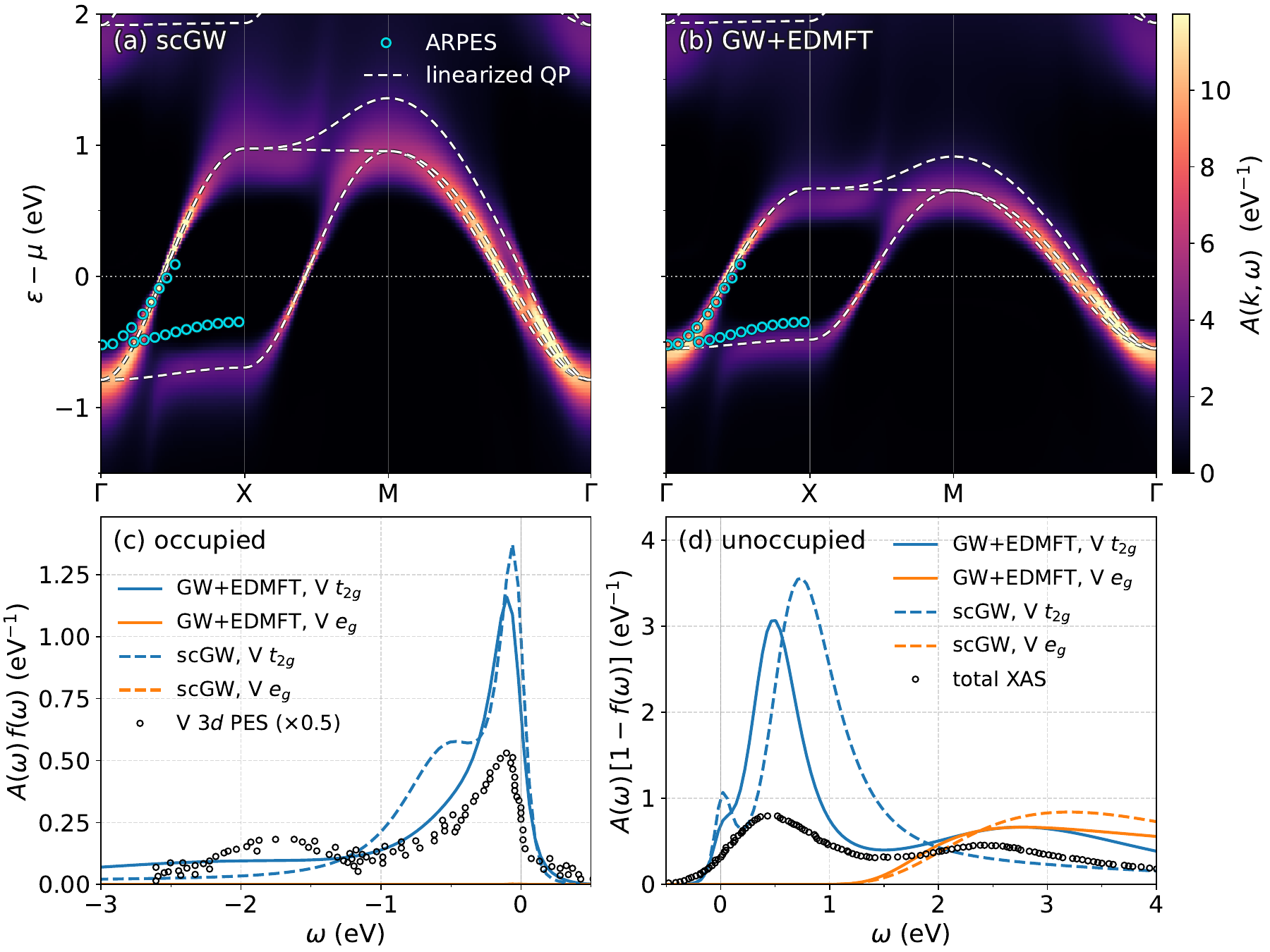} 
    \caption{Spectral functions of SrVO$_3$. (a,b)~Momentum-resolved spectral functions from $\mathrm{sc}GW$ (a) and $GW$+EDMFT (b). White dashed lines show the corresponding linearized quasiparticle bands (Sec.~S2 of the Supplemental Material~\cite{SupplementalMaterial}) and cyan circles the ARPES peak positions of Ref.~\onlinecite{svo_arpes_Takizawa2009}. (c,d)~Local spectral functions of the V-$t_{2g}$ (blue) and V-$e_g$ (orange) orbitals from $GW$+EDMFT (solid) and $\mathrm{sc}GW$ (dashed), multiplied by the Fermi--Dirac function $f(\omega)$ in (c) and by $1-f(\omega)$ in (d), and compared with PES~\cite{svo_pes_Sekiyama2004} (c, scaled by 0.5) and XAS~\cite{svo_pes_xas_Inoue1994} (d) data (black circles).}
    \label{fig:svo_arpes}
    \end{center}
\end{figure}

Figures~\ref{fig:svo_arpes}a,~\ref{fig:svo_arpes}b compare the momentum-resolved spectral functions from $\mathrm{sc}GW$ and $GW$+EDMFT with the ARPES dispersions. The linearized quasiparticle bands (white dashed lines) track the maxima of the interacting spectral functions and facilitate the comparison. Although $\mathrm{sc}GW$ already narrows the $t_{2g}$ bands with respect to PBE, the occupied band bottom at $\Gamma$ remains too deep compared to ARPES. The EDMFT corrections further narrow the $t_{2g}$ manifold on both sides of the Fermi level, and the resulting occupied quasiparticle band follows the ARPES data along $\Gamma$--$X$.
The corresponding quasiparticle weight, evaluated as for LaNiO$_3$~\cite{SupplementalMaterial}, is $Z\simeq0.47$ and nearly momentum-independent along $\Gamma$--$X$. Its inverse, $Z^{-1}\simeq2.1$, is consistent with $m^*/m_b\simeq2$, keeping in mind that, as for LaNiO$_3$, this comparison does not account for the momentum dependence of the self-energy.
A similar value, $Z_{t_{2g}}=0.53$, was reported for the impurity self-energy by the independent fully self-consistent $GW$+EDMFT implementation of Ref.~\onlinecite{GWpEDMFT_KANG2025}, albeit for more localized V-$3d$ orbitals constructed from a $\pm10$~eV window, while multitier $GW$+EDMFT with $d$-only low-energy windows yields $Z_{t_{2g}}=0.6$--$0.65$~\cite{multitier_GWpEDMFT_Boehnke2016,Mushkaev2024}.

The local spectral functions are compared with PES and XAS data in Figs.~\ref{fig:svo_arpes}c,~\ref{fig:svo_arpes}d. On the occupied side, $GW$+EDMFT reproduces the quasiparticle peak and the lineshape of the PES data down to about $-1$~eV, where $\mathrm{sc}GW$ instead overestimates the spectral weight around $-0.5$~eV. Below $-1.5$~eV, $GW$+EDMFT shows only weak, broad spectral weight rather than a pronounced peak at $-1.5$~eV. A weak lower satellite is also found in previous $GW$+EDMFT calculations, both fully self-consistent~\cite{GWpEDMFT_KANG2025} and multitier~\cite{multitier_GWpEDMFT_Boehnke2016,multitier_GWpEDMFT_Nilsson2017,Mushkaev2024}, and recent photoemission experiments have shown that part of the $-1.5$~eV intensity is extrinsic, originating from oxygen-vacancy states~\cite{svo_ovac_Backes2016}.
On the unoccupied side, $GW$+EDMFT reproduces the two-hump structure of the XAS data: a $t_{2g}$ quasiparticle peak at $\simeq0.5$~eV and a broad feature at $2$--$3$~eV. The latter contains both the onset of the $e_g$ manifold and a broad $t_{2g}$ satellite that is absent in $\mathrm{sc}GW$ (dashed line). The spectral function of the local screened interaction $W_{\mathrm{loc}}$ of the V-$t_{2g}$ orbitals exhibits a low-energy pole at $\simeq1.9$~eV, a collective charge excitation of the $t_{2g}$ electrons, which places a plasmonic replica of the quasiparticle peak at $\simeq2.4$~eV, in the energy range of the $t_{2g}$ satellite. This supports the interpretation of this feature as a plasmonic satellite, as proposed on the basis of multitier $GW$+EDMFT~\cite{multitier_GWpEDMFT_Boehnke2016,Mushkaev2024}, while its overlap with the $e_g$ states is consistent with the assignment of Ref.~\onlinecite{one_shoft_GWpEDMFT_Tomczak2014}. Our results indicate that both contribute to the feature observed in XAS.
Altogether, these results confirm that full self-consistency in $G$ and $W$ yields a quantitatively accurate description of a moderately correlated metal, in support of the regime of validity stated in the conclusion.

\end{document}